\documentclass[12pt,amsart,superscriptaddress,amsfonts,floatfix]{iopart}
\usepackage{fullpage,graphicx,epsf}
\usepackage{iopams}%
\usepackage{amsfonts,amssymb}
\usepackage{amscd}
\usepackage{cite}
\usepackage{color}%
\usepackage{graphicx}
\usepackage{epsfig}
\usepackage[english]{babel}
\usepackage{amsfonts}
\usepackage{soul}
\begin{document}

\def\l{\label}
\def\p{\partial}
\def\be{\begin{equation}}
\def\ee{\end{equation}}
\def\bea{\begin{eqnarray}}
\def\eea{\end{eqnarray}}
\def\bef{\begin{figure}}
\def\eef{\end{figure}}
\def\bml{\begin{mathletters}}
\def\eml{\end{mathletters}}
\def\l{\label}
\def\b{\beta}
\def\no{\nonumber}
\def\fr{\frac}
\def\o{\omega}
\def\O{\omega}
\def\p{\partial}
\def\n{\nabla}
\def\a{\alpha}
\def\b{\beta}
\def\eps{\epsilon}
\def\g{\gamma}
\def\d{\delta}
\title{Free energy of a charged oscillator in a magnetic
field and coupled to a heat bath through momentum variables}
\author{Shamik Gupta$^{1,3}$ and Malay Bandyopadhyay$^2$}
\address{$^1$ Laboratoire de Physique de l'\'{E}cole Normale Sup\'{e}rieure de Lyon, Universit\'{e}
de Lyon, CNRS, 46 All\'{e}e d'Italie, 69364 Lyon c\'{e}dex 07, France}
\address{$^2$ Indian Institute of Technology Bhubaneshwar, School of
Basic Sciences, Bhubaneshwar 751007, India}
\address{$^3$ Present address: Laboratoire de Physique Th\'{e}orique et Mod\`{e}les Statistiques, 
UMR 8626, Universit\'{e} Paris-Sud 11 and CNRS, B\^{a}timent 100, Orsay
F-91405, France}
\ead{shamikg1@gmail.com,malay@iitbbs.ac.in}

\begin{abstract}
We obtain an exact formula for the equilibrium free energy of a
charged quantum particle moving in a harmonic potential in the presence
of a uniform external magnetic field and linearly coupled to a heat bath
of independent quantum harmonic oscillators through the momentum variables. We show that the 
free energy has a different expression than that for the
coordinate-coordinate coupling between the particle and the heat-bath
oscillators. For an illustrative heat-bath spectrum, we evaluate the free
energy in the low-temperature limit, thereby showing that the entropy of
the charged particle vanishes at zero temperature, in agreement with the third law of thermodynamics. 
\end{abstract}
\date{\today}
\maketitle
\tableofcontents
\section{Introduction}
Consider the dissipative dynamics of a charged quantum particle moving
in a binding potential in presence of a magnetic field and coupled to the external environment.
Such a situation is frequently encountered in typical experiments
designed to study the magnetic response of a charged particle in
the context of, e.g., Landau diamagnetism
\cite{Vleck:1932}, quantum Hall effect
\cite{Laughlin:1981}, and high-temperature
superconductivity \cite{Ginzburg:1982}. 

In a recent paper \cite{Gupta:2011}, we analyzed the dynamics of the
charged particle by regarding the environment as a quantum mechanical
heat bath or reservoir, and by invoking a gauge-invariant
system-plus-reservoir model. The heat bath was taken to consist of a collection of
independent quantum harmonic
oscillators, while its interaction with the charged particle was
modelled in terms of bilinear coupling between the momentum variables of the particle and the
oscillators. We considered a harmonic binding potential and a magnetic
field which is uniform in space. The Hamiltonian of the system is
\bea
H_0=\fr{1}{2m}\Big({\bf p}-\fr{e}{c}{\bf A}\Big)^2+\fr{1}{2}m\o_0^2{\bf r}^2+\sum_{j=1}^N\Big[\fr{1}{2m_j}\Big({\bf p}_j-g_j{\bf
p}+\fr{g_je}{c}{\bf A}\Big)^2+\fr{1}{2}m_j\o_j^2{\bf q}_j^2\Big],
\nonumber \\ 
\l{H}
\eea
where
$e,m,{\bf p},{\bf r}$ are, respectively, the charge, the mass, the
momentum operator and the coordinate operator of the particle, while
$\o_0$ is the frequency characterizing its motion in the harmonic
potential. The $j$th heat-bath oscillator has mass $m_j$, frequency
$\o_j$, coordinate operator ${\bf q}_j$, and momentum operator ${\bf
p}_j$. The dimensionless parameter $g_j$ describes the coupling between
the particle and the $j$th oscillator. The speed of light in vacuum is
denoted by $c$. The vector potential ${\bf A}={\bf A}({\bf r})$ is
related to the uniform external magnetic field ${\bf B}=(B_x,B_y,B_z)$ through ${\bf
B}=\n \times {\bf A}({\bf r})$. The field has the magnitude
$B=\sqrt{B_x^2+B_y^2+B_z^2}$. 
The commutation relations for
the different coordinate and momentum operators are
\be
[r_\a,p_\b]=i\hbar \delta_{\a \b}, [q_{j\a},p_{k\b}]=i\hbar \delta_{jk}\delta_{\a \b},
\l{canonical-commutation}
\ee
while all other commutators vanish. In the above equation, $\delta_{jk}$
denotes the Kronecker Delta function. Throughout this paper, Greek indices ($\a, \b, \ldots$) refer to the three spatial directions, while Roman indices ($i,j,k,\ldots$) represent the heat-bath oscillators.
Moreover, we use Einstein summation convention for the Greek indices. 
Let us remark that momentum-momentum coupling has been considered earlier in the literature
\cite{Leggett:1984}, and, in particular, to model the physical situation
of a single Josephson junction interacting with the blackbody
electromagnetic field in the dipole approximation
\cite{Cuccoli:2001,Ankerhold:2007}. Our model Hamiltonian is similar to
that considered in Refs. \cite{Cuccoli:2001,Ankerhold:2007}, the additional interesting feature
that we consider here is the inclusion of the effects of an external magnetic field. 

In Ref. \cite{Gupta:2011}, we derived a quantum Langevin equation (QLE)
satisfied by the coordinate operator of the charged particle. In this
equation, coupling to the bath is described solely by (i) an operator-valued
random force, and (ii) a mean force characterized by a memory function.
We showed that similar to the case of coordinate-coordinate coupling
between the charged particle and the heat-bath oscillators, the QLE involves a quantum-generalized Lorentz force term, and also
that the random force does not depend on the magnetic field. This latter
force, nevertheless, has a modified form, with symmetric correlation and
unequal time commutator different from their corresponding expressions in the case of coordinate-coordinate coupling. Other differences include (i) the memory function has an explicit dependence on the magnetic field, and (ii) the inertial term and the harmonic potential term in the QLE get renormalized by the coupling constants.

In this paper, we further our study of the system (\ref{H}) by deriving
an exact formula for the mean internal energy and thence, the free energy of the
charged particle in thermal equilibrium. Knowing either the mean internal energy
or the free energy, one can compute by taking suitable
derivatives a host of thermodynamic
functions, e.g., specific heat, susceptibility,
etc \cite{Ford:2007,Ingold:2009,Dattagupta:2010,Ingold:2012}. Our result shows important differences in the form of the 
free energy with respect to that for the coordinate-coordinate coupling
\cite{Li:1990}. For an illustrative heat-bath spectrum, we evaluate the
free energy in the low-temperature limit, thereby showing that the entropy of the charged particle vanishes
at zero temperature, in conformity with the third law of thermodynamics. 
 
The paper is structured as follows. In the following section, we write down the equations of motion for the charged
particle and the heat-bath oscillators. In section
\ref{mean-energy}, we
present a detailed derivation of the equilibrium free energy of the charged
particle, and point out its differences from the case of
coordinate-coordinate coupling. In the section that follows, we consider
an example of a heat-bath spectrum and analyze the free energy of the charged particle in the limit of
low temperatures. We draw our
conclusions in section \ref{conclusions}.

\section{Equations of motion}
For the system (\ref{H}), the Heisenberg equations of motion for the charged particle are 
\bea
\dot{\bf r}&=&\fr{1}{i\hbar}[{\bf r},H_0]=\fr{1}{m}\Big({\bf p}-\fr{e}{c}{\bf A}\Big)-\sum_{j=1}^N
\fr{g_j}{m_j}\Big({\bf p}_j-g_j{\bf p}+\fr{g_je}{c}{\bf A}\Big),\l{qdot}
\eea
and
\bea
\dot{\bf p}&=&\fr{1}{i\hbar}[{\bf p},H_0]=\fr{e}{c}(\dot{\bf r}\times {\bf B})+\fr{e}{c}(\dot{\bf r}\cdot\n){\bf
A}+\fr{i\hbar e}{2m_{\rm r}c}\n (\n\cdot{\bf A})-m\o_0^2 {\bf r},
\l{pdot}
\eea
where $m_{\rm r}$ is the ``renormalized mass",
\be
m_{\rm r}\equiv m/\Big[1+\sum_{j=1}^N \fr{g_j^2m}{m_j}\Big], 
\l{mrdefinition}
\ee
and dots denote differentiation with respect to time. 
The Heisenberg equations for the heat-bath oscillators are
\bea
\dot{\bf q}_j&=&\fr{1}{i\hbar}[{\bf q}_j,H_0]=\fr{1}{m_j}\Big({\bf p}_j-g_j{\bf p}+\fr{g_je}{c}{\bf
A}\Big), \l{qjdot} 
\eea
and
\bea
\dot{\bf p}_j&=&\fr{1}{i\hbar}[{\bf p}_j,H_0]=-m_j\o_j^2{\bf q}_j.
\l{pjdot}
\eea
These equations lead to
\be
m_{\rm r}\ddot {\bf r}=-m\o_0^2{\bf r}+\fr{e}{c}(\dot{\bf r}\times {\bf
B})+\sum_{j=1}^N g_jm_{\rm r}\o_j^2{\bf q}_j,
\l{particle-EOM}
\ee
and
\be
m_j\ddot{\bf q}_j=-m_j\o_j^2{\bf q}_j+g_jm\o_0^2{\bf r}-\fr{g_j
e}{c}(\dot{\bf r} \times {\bf B}).
\l{heat-bath-EOM}
\ee
For details on the derivation of Eqs. (\ref{qdot}) -
(\ref{heat-bath-EOM}), see Ref. \cite{Gupta:2011}. 
 \section{Equilibrium free energy}
\l{mean-energy}
In this section, we present our computation of the equilibrium free
energy of the charged particle. The steps of computation are as follows:
We assume that the system (\ref{H}) is in {\em weak} contact with a {\em
super}-bath that allows the system to attain thermal equilibrium at
temperature $T$. In
such an equilibrium state, we obtain the mean internal 
energy $U_0(T,B)$ of the particle, defined as the equilibrium mean
internal energy of the
system of the charged particle interacting with the heat bath, given by
$\langle H_0 \rangle$, minus that of the heat bath in the absence of
coupling with the particle. In our case of a heat bath comprising
independent quantum harmonic oscillators, this latter quantity is given
by
\be
U_B(T)=\sum_{j=1}^N \fr{3\hbar
\o_j}{2}\coth\Big[\fr{\hbar\o_j}{2k_BT}\Big].
\ee
Once $U_0(T,B)$ has been computed,
the free energy $F_0(T,B)$ of the charged particle is obtained by using
the usual relationship between the two: $U_0(T,B)=F_0(T,B)-T\fr{\partial
F_0(T,B)}{\partial T}$. 

Let us start with computing $\langle H_0 \rangle$. Using Eqs.~(\ref{qdot})
and (\ref{qjdot}), we get
\bea
\langle H_0 \rangle &=&\Big[\fr{1}{2}m\langle \dot{\bf r}^2 \rangle +
\fr{1}{2}m\o_0^2 \langle {\bf r}^2 \rangle \Big]+\fr{m}{2}\Big[\sum_{j=1}^N g_j \langle \dot{\bf
r}\cdot\dot{\bf q}_j+ \dot{\bf q}_j\cdot\dot{\bf
r}\rangle\nonumber \\
&+&\sum_{j,k=1}^Ng_jg_k \langle \dot{\bf q}_j\cdot\dot{\bf
q}_k \rangle\Big]+\sum_{j=1}^N\Big[\fr{1}{2}m_j \langle \dot{\bf q}_j^2 \rangle + \fr{1}{2}m_j\o_j^2 \langle {\bf
q}_j^2 \rangle\Big].
\l{H0-average}
\eea
Then, to find $\langle H_0 \rangle$, we need equilibrium averages, such as $\langle {\bf r}^2 \rangle$ and $\langle \dot{\bf
r}^2 \rangle$, which may be obtained by knowing the equilibrium
autocorrelation function of the position of the charged
particle, defined as
\bea
\psi^{rr}_{\rho\sigma}(t-t') &\equiv& \fr{1}{2}\langle
r_\rho(t)r_\sigma(t')+r_\sigma(t')r_\rho(t)\rangle.
\l{psirhosigmat}
\eea
Similarly, $\langle {\bf q}_j^2 \rangle$, $\langle \dot{\bf q}_j^2
\rangle$, and $\langle \dot{\bf q}_j\cdot\dot{\bf q}_k \rangle$ are obtained
from the position autocorrelation function of the heat-bath oscillators, defined as
\bea
\psi^{qq}_{jk,\rho\sigma}(t-t') &\equiv& \fr{1}{2}\langle
q_{j\rho}(t)q_{k\sigma}(t')+q_{k\sigma}(t')q_{j\rho}(t)\rangle,
\l{psijkrhosigmat}
\eea
while $\langle \dot{\bf r} \cdot \dot{\bf q}_j + \dot{\bf q}_j \cdot \dot{\bf r}\rangle$ may be computed from the correlations
\bea
\psi^{rq}_{j,\rho\sigma}(t-t') &\equiv& \fr{1}{2}\langle
r_{\rho}(t)q_{j\sigma}(t')+q_{j\sigma}(t')r_{\rho}(t)\rangle,
\l{psijrhosigmat}
\eea
and
\bea
\psi^{qr}_{j,\rho\sigma}(t-t') &\equiv& \fr{1}{2}\langle
q_{j,\rho}(t)r_{\sigma}(t')+r_{\sigma}(t')q_{j,\rho}(t)\rangle.
\l{psijrhosigmat1}
\eea

Correlations such as in Eqs. (\ref{psirhosigmat}) - (\ref{psijrhosigmat1}) are computed conveniently by invoking the
fluctuation-dissipation theorem that relates such equilibrium
correlations to response of the system to small external
perturbations \cite{Kubo:1966}. To this end, consider a weak external force ${\bf f}(t)$ to act on the charged
particle and another set of weak external forces $\{{\bf f}_j(t), j=1,2,\ldots,N\}$ to
act on the oscillators. Here, ${\bf f}(t)$ and ${\bf f}_j(t)$ are
$c$-number functions of time. We take the perturbed Hamiltonian to be of the form
\be
H=H_0-{\bf r}\cdot {\bf f}(t)-\sum_{j=1}^N {\bf q}_j \cdot {\bf f}_j(t).
\ee
In the presence of these external forces, the Heisenberg equations of motion (\ref{qdot}) and (\ref{qjdot}) remain
the same, while those given by Eqs. (\ref{pdot}) and (\ref{pjdot}),
respectively, are
modified to 
\bea
\dot{\bf p}=\fr{e}{c}(\dot{\bf r}\times {\bf B})+\fr{e}{c}(\dot{\bf r}\cdot\n){\bf
A}+\fr{i\hbar e}{2m_{\rm r}c}\n (\n\cdot{\bf A})-m\o_0^2 {\bf r}+{\bf f},
\l{pdotagain}
\eea
and
\be
\dot{\bf p}_j=-m_j\o_j^2{\bf q}_j+{\bf f}_j.
\l{pjdotagain}
\ee
Consequently, Eqs. (\ref{particle-EOM}) and (\ref{heat-bath-EOM}),
respectively, are modified to
\be
m_{\rm r}\ddot {\bf r}=-m\o_0^2{\bf r}+\fr{e}{c}(\dot{\bf r}\times {\bf
B})+\sum_{j=1}^N g_jm_{\rm r}\o_j^2{\bf q}_j-\sum_{j=1}^N\fr{g_jm_{\rm
r}}{m_j}{\bf f}_j+{\bf f},
\l{particle-EOM-extforce}
\ee
and
\be
m_j\ddot{\bf q}_j=-m_j\o_j^2{\bf q}_j+g_jm\o_0^2{\bf r}-\fr{g_j
e}{c}(\dot{\bf r} \times {\bf B})+{\bf f}_j-g_j{\bf f}.
\l{heat-bath-EOM-extforce}
\ee

Taking the Fourier transform of Eqs. (\ref{particle-EOM-extforce}) and
(\ref{heat-bath-EOM-extforce}), we get  
\bea
&&\Big[\delta_{\rho\sigma}(-m_{\rm r}\O^2+m\o_0^2)+\fr{i\O
e}{c}\eps_{\rho\sigma\eta}B_\eta\Big]\bar{r}_\sigma-\sum_{j=1}^N
g_jm_{\rm r}\o_j^2\bar{q}_{j\rho}=-\sum_{j=1}^N\fr{g_jm_{\rm
r}}{m_j}\bar{f}_{j\rho}+\bar{f}_\rho, \nonumber \\
\l{particle-EOM-FT}
\eea
and 
\bea
&&(-m_j\O^2+m_j\o_j^2)\bar{q}_{j\rho}-\Big(g_jm\o_0^2\delta_{\rho\sigma}+\fr{i\O
g_je}{c}\eps_{\rho\sigma\eta}B_\eta\Big)\bar{r}_\sigma=\bar{f}_{j\rho}-g_j\bar{f}_{\rho},
\l{heat-bath-EOM-FT}
\eea
where $\epsilon_{\rho\sigma\eta}$ is the Levi-Civita symbol.
Equations (\ref{particle-EOM-FT}) and (\ref{heat-bath-EOM-FT}) give
\be
D_{\rho\sigma}(\O)\bar{r}_\sigma=G(\O)\bar{f}_\rho+\sum_{j=1}^N\fr{g_jm_{\rm
r}\O^2}{m_j(\o_j^2-\O^2)}\bar{f}_{j\rho},
\l{rsigmaeq0}
\ee
where
\be
G(\O)=1-\sum_{j=1}^N\fr{(g_j)^2 m_{\rm
r}\o_j^2}{m_j(\o_j^2-\O^2)},
\l{Gomega-defn}
\ee
\be
D_{\rho\sigma}(\O)=\lambda(\O)\delta_{\rho\sigma}+\fr{i\O
eG(\O)}{c}\epsilon_{\rho\sigma\eta}B_\eta,
\ee
and
\be
\lambda(\O)=-m_{\rm r}\O^2+m\o_0^2G(\O).
\l{lambda}
\ee

From Eq. (\ref{rsigmaeq0}), we get 
\be
\bar{r}_\rho=G(\O)\alpha_{\rho\gamma}(\O)\bar{f}_\gamma+\sum_{j=1}^N\beta_{j,\rho\gamma}(\O)\bar{f}_{j\gamma},
\l{requation}
\ee
where
\bea
\alpha_{\rho\gamma}(\O)&=&[D(\O)^{-1}]_{\rho\gamma}\nonumber \\
&=&\Big[(\lambda(\O))^2 \delta_{\rho\gamma} -
\Big(\fr{\O eG(\O)}{c}\Big)^2B_\rho
B_\gamma-\fr{i\O\lambda(\O)eG(\O)}{c}\epsilon_{\rho\gamma\eta}B_{\eta}\Big]/{\rm Det}~D(\O),
\nonumber \\
\l{alphacompact}
\eea
\be
{\rm Det}~D(\O)=\lambda(\O)\Big[(\lambda(\O))^2-\Big(\fr{\O B
eG(\O)}{c}\Big)^2\Big],
\l{DetDomega}
\ee
and
\be
\beta_{j,\rho\gamma}(\O)=\fr{g_j m_{\rm
r}\O^2}{m_j(\o_j^2-\O^2)}\alpha_{\rho\gamma}(\O).
\l{betadefinition}
\ee

Using Eq. (\ref{requation}) in either Eq. (\ref{particle-EOM-FT})
or Eq. (\ref{heat-bath-EOM-FT}) gives 
\be
\bar{q}_{j\rho}=\Delta_{j,\rho\gamma}(\O)\bar{f}_\gamma+\sum_{k=1}^N
\gamma_{jk,\rho\gamma}(\O)\bar{f}_{k\gamma},
\l{qjequation}
\ee
where
\bea
\Delta_{j,\rho\gamma}(\O)&=&\fr{g_j m G(\O)}{m_j(\o_j^2-\O^2)}\Big(\o_0^2\delta_{\rho\sigma}+\fr{i\O
e}{mc}\eps_{\rho\sigma\eta}B_\eta\Big)\alpha_{\sigma\gamma}(\O)-\fr{g_j}{m_j(\o_j^2-\O^2)}\delta_{\rho\gamma},
\l{deltadefinition}
\eea
and
\bea
\gamma_{jk,\rho\gamma}(\O)&=&\fr{g_jg_kmm_{\rm r}\O^2}{m_j m_k(\o_j^2-\O^2)(\o_k^2-\O^2)}\Big(\o_0^2\delta_{\rho\sigma}+\fr{i\O
e}{mc}\eps_{\rho\sigma\eta}B_\eta\Big)\alpha_{\sigma\gamma}(\O)\nonumber
\\
&+&\fr{\delta_{jk}\delta_{\rho\gamma}}{m_k(\o_k^2-\O^2)}.
\l{gammadefinition}
\eea

The functions $\alpha_{\rho\gamma}(\O), \beta_{j,\rho\gamma}(\O),
\Delta_{j,\rho\gamma}(\O)$, and $\gamma_{jk,\rho\gamma}(\O)$ are coefficient
matrices of the response of the system to the external perturbations, ${\bf
f}(t)$ and ${\bf f}_j(t)$, and may be interpreted as generalized susceptibilities.

We may now use the fluctuation-dissipation theorem which relates the
Fourier transform $\bar{\psi}^{rr}_{\rho\sigma}(\O)$ of the position autocorrelation
function $\psi^{rr}_{\rho\sigma}(t-t')$ to 
$G(\O)\alpha_{\rho\sigma}(\O)$:
\bea
\bar{\psi}^{rr}_{\rho\sigma}(\O)&=&\fr{\hbar}{2i}\coth\Big[\fr{\hbar\O}{2k_BT}\Big][G(\O)\alpha_{\rho\sigma}(\O)-G^*(\O)\alpha^*_{\sigma\rho}(\O)],
\l{psibarrhosigma11}
\eea
where $*$ denotes complex conjugation. Using Eq. (\ref{Gomega-defn}), we
see that $G(\O)$ is real, so that 
\bea
\bar{\psi}^{rr}_{\rho\sigma}(\O)&=&\fr{\hbar}{2i}\coth\Big[\fr{\hbar\O}{2k_BT}\Big]G(\O)[\alpha_{\rho\sigma}(\O)-\alpha^*_{\sigma\rho}(\O)].
\l{psibarrhosigma1}
\eea
Similarly, one has
\bea
\bar{\psi}^{qq}_{jk,\rho\sigma}(\O)&=&\fr{\hbar}{2i}\coth\Big[\fr{\hbar\O}{2k_BT}\Big][\gamma_{jk,\rho\sigma}(\O)-\gamma^*_{jk,\sigma\rho}(\O)],
\l{psibarrhosigma2}
\eea
\be
\bar{\psi}^{rq}_{j,\rho\sigma}(\O)=\fr{\hbar}{2i}\coth\Big[\fr{\hbar\O}{2k_BT}\Big][\beta_{j,\rho\sigma}(\O)-\beta^*_{j,\sigma\rho}(\O)],
\l{psibarrhosigma3}
\ee
and
\be
\bar{\psi}^{qr}_{j,\rho\sigma}(\O)=\fr{\hbar}{2i}\coth\Big[\fr{\hbar\O}{2k_BT}\Big][\Delta_{j,\rho\sigma}(\O)-\Delta^*_{j,\sigma\rho}(\O)].
\l{psibarrhosigma4}
\ee

In passing, we note that 
\be
\alpha^*_{\sigma\rho}(\O)=\alpha_{\sigma\rho}(-\O),
\ee
which implies that ${\rm Im}[\alpha_{\sigma\rho}(\O)]$ is an odd function of
$\O$, while ${\rm Re}[\alpha_{\sigma\rho}(\O)]$ is an even function of
$\O$. These properties are also shared by $\beta_{j,\sigma\rho}(\O),
\Delta_{j,\sigma\rho}(\O)$, and $\gamma_{jk,\sigma\rho}(\O)$.

In the following, we use Eqs. (\ref{psibarrhosigma1}) -
(\ref{psibarrhosigma4}) to evaluate the various averages involved in
computing $\langle H_0 \rangle$. 

\subsection{Computation of various averages}
\subsubsection{Averages involving ${\bf r}$:}
We start with decomposing $\alpha_{\rho\sigma}$ into symmetric and
antisymmetric parts as
\be
\alpha_{\rho\sigma}(\O)=\alpha^{\rm s}_{\rho\sigma}(\O)+\alpha^{\rm a}_{\rho\sigma}(\O).
\ee
We then have
\bea
\alpha_{\rho\sigma}(\O)-\alpha^*_{\sigma\rho}(\O)&=&[\alpha^{\rm s}_{\rho\sigma}(\O)-\alpha^{\rm
s}_{\rho\sigma}(\O)^*]+[\alpha^{\rm a}_{\rho\sigma}(\O)+\alpha^{\rm a}_{\rho\sigma}(\O)^*]\nonumber
\\
&=&2i{\rm Im}[\alpha^{\rm s}_{\rho\sigma}(\O)]+2{\rm Re}[\alpha^{\rm a}_{\rho\sigma}(\O)].
\l{alphasymmetry}
\eea
Using Eqs. (\ref{psirhosigmat}), (\ref{psibarrhosigma1}),
(\ref{alphasymmetry}), and the properties that ${\rm Im}[\alpha_{\rho\sigma}(\O)]$ is an odd function of
$\O$, while both $G(\O)$ and ${\rm Re}[\alpha_{\rho\sigma}(\O)]$ are
even functions of
$\O$, we get 
\bea
&&\fr{1}{2}\langle r_\rho(t)r_\sigma(t')+r_\sigma(t')r_\rho(t)\rangle\nonumber \\
&&=\fr{\hbar}{\pi}\int_0^\infty d\O~
G(\O){\rm Im}[\alpha^{\rm
s}_{\rho\sigma}(\O)]\coth\Big[\fr{\hbar\O}{2k_BT}\Big]\cos[\O(t-t')]\nonumber
\\
&&-\fr{\hbar}{\pi}\int_0^\infty d\O~
G(\O){\rm Re}[\alpha^{\rm a}_{\rho\sigma}(\O)]\coth\Big[\fr{\hbar\O}{2k_BT}\Big]\sin[\O(t-t')], 
\l{particlecorrelation}
\eea
which for $\rho=\sigma$ gives
\bea
&&\fr{1}{2}\langle {\bf r}(t)\cdot {\bf r}(t')+{\bf r}(t')\cdot {\bf
r}(t)\rangle\nonumber \\
&&=\fr{\hbar}{\pi}\int_0^\infty
d\O~
G(\O){\rm Im}[\alpha_{\rho\rho}(\O)]\coth\Big[\fr{\hbar\O}{2k_BT}\Big]\cos[\O(t-t')].
\l{particlecorrelation1}
\eea
Then, putting $t=t'$, we get
\be
\langle {\bf r}^2\rangle=\fr{\hbar}{\pi}\int_0^\infty d\O~
G(\O){\rm Im}[\alpha_{\rho\rho}(\O)]\coth\Big[\fr{\hbar\O}{2k_BT}\Big],
\l{rsqavg}
\ee
where from Eq. (\ref{alphacompact}), we have 
\bea
&&\alpha_{\rho\rho}(\O)=\Big[(\lambda(\O))^2 \delta_{\rho\rho} -
\Big(\fr{\O eG(\O)}{c}\Big)^2B_\rho
B_\rho\Big]/{\rm Det}~D(\O).
\l{alphasymmetric}
\eea
Differentiating Eq. (\ref{particlecorrelation1}) successively with
respect to $t$ and $t'$, and finally putting $t=t'$, we get
\bea
\langle \dot{\bf r}^2\rangle
=\fr{\hbar}{\pi}\int_0^\infty d\O~
G(\O){\rm Im}[\alpha_{\rho\rho}(\O)]\coth\Big[\fr{\hbar\O}{2k_BT}\Big]\O^2.
\l{rdotsqavg}
\eea
\subsubsection{Averages involving ${\bf q}_j$:}
One may proceed similarly to the preceding subsection to obtain
\bea
&&\fr{1}{2}\langle
{\bf q}_{j}(t)\cdot{\bf q}_{k}(t')+{\bf q}_{k}(t')\cdot{\bf q}_{j}(t)\rangle\nonumber
\\
&&=\fr{\hbar}{\pi}\int_0^\infty d\O~ {\rm Im}[\gamma_{jk,
\rho\rho}(\O)]\coth\Big[\fr{\hbar\O}{2k_BT}\Big]\cos[\O(t-t')],
\l{qjqk}
\eea
where $\gamma_{jk,\rho\rho}(\O)$ is obtained from Eq.
(\ref{gammadefinition}) as
\bea
\gamma_{jk,\rho\rho}(\O)
&=&\fr{g_j g_k m m_{\rm r}\O^2\o_0^2}{m_j
m_k(\o_j^2-\O^2)(\o_k^2-\O^2)}\alpha_{\rho\rho}(\O)\nonumber \\
&-&\fr{g_j g_k m_{\rm r}\O^4\lambda(\O)e^2G(\O)}{m_j
m_k(\o_j^2-\O^2)(\o_k^2-\O^2)c^2{\rm Det}~D(\O)}(\delta_{\rho\rho}B^2-B_\rho
B_\rho)\nonumber \\
&+&\fr{\delta_{jk}\delta_{\rho\rho}}{m_k(\o_k^2-\O^2)}. 
\l{gammasymmetric}
\eea
One therefore gets
\be
\langle {\bf q}_j^2\rangle=\fr{\hbar}{\pi}\int_0^\infty d\O~ {\rm Im}[\gamma_{jj,\rho\rho}(\O)]\coth\Big[\fr{\hbar\O}{2k_BT}\Big],
\l{qjsqavg}
\ee
\be
\langle \dot{\bf q}_j^2\rangle=\fr{\hbar}{\pi}\int_0^\infty
d\O~
{\rm Im}[\gamma_{jj,\rho\rho}(\O)]\coth\Big[\fr{\hbar\O}{2k_BT}\Big]\O^2,
\l{qjdotsqavg}
\ee
and
\bea
&&\langle{\bf \dot{q}}_{j}\cdot{\bf
\dot{q}}_{k}\rangle=\fr{\hbar}{\pi}\int_0^\infty d\O~ {\rm Im}[\gamma_{jk,
\rho\rho}(\O)]\coth\Big[\fr{\hbar\O}{2k_BT}\Big]\O^2. 
\l{qjqk1}
\eea
\subsubsection{Averages involving ${\bf r}$ and ${\bf q}_j$:}
We have
\be
\fr{1}{2}\langle
r_\rho(t)q_{j\sigma}(t')+q_{j\sigma}(t')r_\rho(t)\rangle=\fr{1}{2\pi}\int_{-\infty}^\infty
d\O~ e^{-i\O (t-t')}\bar{\psi}^{rq}_{j,\rho\sigma}(\O),
\l{psirhosigmatagain}
\ee
and
\be
\fr{1}{2}\langle
q_{j\sigma}(t')r_\rho(t)+r_{\rho}(t)q_{j\sigma}(t')\rangle=\fr{1}{2\pi}\int_{-\infty}^\infty
d\O~ e^{-i\O (t'-t)}\bar{\psi}^{qr}_{j,\sigma\rho}(\O).
\l{psirhosigmataagain3}
\ee
Adding Eq. (\ref{psirhosigmataagain3}) to Eq. (\ref{psirhosigmatagain}), we get
\bea
&&\langle
r_{\rho}(t)q_{j\sigma}(t')+q_{j\sigma}(t')r_\rho(t)\rangle=\fr{1}{2\pi}\int_{-\infty}^\infty d\O~ \Big[e^{-i\O
(t-t')}\bar{\psi}^{rq}_{j,\rho\sigma}(\O)+e^{-i\O
(t'-t)}\bar{\psi}^{qr}_{j,\sigma\rho}(\O)\Big].\nonumber \\  
\l{psibarrhosigma5}
\eea
Differentiating Eq. (\ref{psibarrhosigma5}) successively with respect to
$t$ and $t'$, and then putting $t=t'$, we get
\bea
\langle
\dot{r}_{\rho}\dot{q}_{j\sigma}+\dot{q}_{j\sigma}\dot{r}_\rho\rangle=\fr{1}{2\pi}\int_{-\infty}^\infty d\O~ \O^2
\Big[\bar{\psi}^{rq}_{j,\rho\sigma}(\O)+\bar{\psi}^{qr}_{j,\sigma\rho}(\O)\Big].
\l{psibarrhosigma8}
\eea
Using Eqs. (\ref{psibarrhosigma3}) and (\ref{psibarrhosigma4}) for
$\bar{\psi}^{rq}_{j,\rho\sigma}(\O)$ and
$\bar{\psi}^{qr}_{j,\sigma\rho}(\O)$, respectively, and noting that the
imaginary part of both $\beta_{j,\rho\sigma}(\O)$ and
$\Delta_{j,\rho\sigma}(\O)$ is an odd function of
$\O$, while their real parts are even functions of
$\O$, we get
\bea
&&\langle \dot{\bf q}_{j}\cdot\dot{\bf r}+\dot{\bf
r}\cdot\dot{\bf q}_{j}\rangle\nonumber \\
&&=\fr{\hbar}{\pi}\int_0^\infty
d\O~\coth\Big[\fr{\hbar\O}{2k_BT}\Big] 
\O^2\Big[{\rm Im}[\beta_{j,\rho\rho}(\O)]+{\rm Im}[\Delta_{j,\rho\rho}(\O)]\Big], 
\l{qjdotrdotavg}
\eea
where from Eq. (\ref{deltadefinition}), we get
\bea
\Delta_{j,\rho\rho}(\O)
&=&\fr{g_jmG(\O)\o_0^2}{m_j(\o_j^2-\O^2)}\alpha_{\rho\rho}(\O)
-\fr{g_j(G(\O))^2\lambda(\O)\O^2e^2}{m_j(\o_j^2-\O^2)c^2{\rm Det}~D(\O)}(\delta_{\rho\rho}
B^2-B_\rho B_\rho)\nonumber \\
&-&\fr{g_j}{m_j(\o_j^2-\O^2)}\delta_{\rho\rho}.
\label{deltasymmetric}
\eea
\subsection{Mean internal energy}
Using Eqs. (\ref{rsqavg}), (\ref{rdotsqavg}), (\ref{qjsqavg}),
(\ref{qjdotsqavg}), (\ref{qjqk1}), and (\ref{qjdotrdotavg}) in Eq. (\ref{H0-average}), we get
\bea
\langle H_0 \rangle&=&\fr{\hbar}{\pi}\int_0^\infty d\O
\coth\Big[\fr{\hbar\O}{2k_BT}\Big]\Big[\fr{1}{2}m(\O^2+\o_0^2)G(\O){\rm Im}[\alpha_{\rho\rho}(\O)]\nonumber
\\
&+&\sum_{j=1}^N\fr{mg_j\O^2}{2}\Big({\rm
Im}[\beta_{j,\rho\rho}(\O)]+{\rm
Im}[\Delta_{j,\rho\rho}(\O)]\Big)\nonumber \\
&+&\sum_{j,k=1}^N\fr{mg_jg_k\O^2}{2}{\rm Im}[\gamma_{jk,
\rho\rho}(\O)]\nonumber \\
&+&\sum_{j=1}^N\fr{1}{2}m_j(\O^2+\o_j^2){\rm Im}[\gamma_{jj,\rho\rho}(\O)]\Big].
\eea

The mean internal energy of the heat bath is
\bea
\langle H_B \rangle&=&\sum_{j=1}^N \fr{1}{2}m_j\o_j^2\langle {\bf q}_j^2
\rangle +\sum_{j=1}^N\fr{1}{2}m_j\langle \dot{\bf q}_j^2
\rangle\nonumber \\
&=&\sum_{j=1}^N\fr{m_j\hbar}{2\pi}\int_0^\infty
d\O\coth\Big[\fr{\hbar\O}{2k_BT}\Big]
{\rm Im}[\gamma_{jj,\rho\rho}(\O)](\O^2+\o_j^2) \nonumber \\
&=&\sum_{j=1}^N \fr{3\hbar}{2\pi}\int_0^\infty
d\O\coth\Big[\fr{\hbar\O}{2k_BT}\Big]\Big(\fr{\o^2+\o_j^2}{\o_j^2-\o^2}\Big)
\nonumber \\
&+&\fr{\hbar}{2\pi}\int_0^\infty d\O
\coth\Big[\fr{\hbar\O}{2k_BT}\Big]\sum_{j=1}^Nm_j(\O^2+\o_j^2)\nonumber
\\
&\times&{\rm Im}\Big[\fr{(g_j)^2 m
m_{\rm r}\O^2\o_0^2}{(m_j)^2(\o_j^2-\O^2)^2}\alpha_{\rho\rho}(\O)-\fr{2(g_j)^2 
m_{\rm r}\O^4\lambda(\O)
B^2e^2G(\O)}{(m_j)^2(\o_j^2-\O^2)^2c^2{\rm Det}~D(\O)}\Big] \l{secondline} \\
&=&\sum_{j=1}^N \fr{3\hbar
\o_j}{2}\coth\Big[\fr{\hbar\o_j}{2k_BT}\Big]\nonumber \\
&+&\fr{\hbar}{2\pi}\int_0^\infty d\O
\coth\Big[\fr{\hbar\O}{2k_BT}\Big]\sum_{j=1}^Nm_j(\O^2+\o_j^2)\nonumber
\\
&\times&{\rm Im}\Big[\fr{(g_j)^2 m
m_{\rm r}\O^2\o_0^2}{(m_j)^2(\o_j^2-\O^2)^2}\alpha_{\rho\rho}(\O)-\fr{2(g_j)^2 
m_{\rm r}\O^4\lambda(\O)
B^2e^2G(\O)}{(m_j)^2(\o_j^2-\O^2)^2c^2{\rm Det}~D(\O)}\Big].
\l{HBaverage}
\eea
Here, in obtaining Eq. (\ref{secondline}), we have used Eq.
(\ref{gammasymmetric}).
To arrive at Eq. (\ref{HBaverage}), we have performed the first
integral on the right hand side of Eq. (\ref{secondline}) by noting that
$\o$ in the integral is approached from above the real axis, $\o
\rightarrow \o+i0^+$, and by using the result that 
\be
\fr{1}{\O-\o_j+i0^+}={\rm P}\Big[\fr{1}{\O-\o_j}\Big]-i\pi\delta(\O-\o_j),
\ee
with P denoting the principal value.
Note that the first term on the right hand side of Eq. (\ref{HBaverage})
is the mean internal energy $U_B(T)$ of the heat bath in the absence of any coupling with
the charged particle:
\be
U_B(T)=\sum_{j=1}^N \fr{3\hbar
\o_j}{2}\coth\Big[\fr{\hbar\o_j}{2k_BT}\Big].
\ee

We now obtain the mean internal energy $U_0(T,B)$ of the particle as the
mean internal energy of the
system of the charged particle interacting with the heat bath minus the mean
internal energy of the heat bath in the absence of coupling with the particle:
\bea
\hspace{-1.25cm}U_0(T,B)&=&\langle H_0 \rangle - U_B(T)\nonumber \\
&=&\fr{\hbar}{\pi}\int_0^\infty d\O
\coth\Big[\fr{\hbar\O}{2k_BT}\Big]\Big[\fr{1}{2}m(\O^2+\o_0^2)G(\O){\rm
Im}[\alpha_{\rho\rho}(\O)]\nonumber \\
&+&\sum_{j=1}^N\fr{mg_j\O^2}{2}\Big({\rm Im}\Big[\fr{g_j m_{\rm
r}\O^2}{m_j(\o_j^2-\O^2)}\alpha_{\rho\rho}(\O)\Big]\nonumber \\
&+&{\rm
Im}\Big[\fr{g_jmG(\O)\o_0^2}{m_j(\o_j^2-\O^2)}\alpha_{\rho\rho}(\O)-\fr{2g_j(G(\O))^2\lambda(\O)\O^2e^2B^2}{m_j(\o_j^2-\O^2)c^2{\rm
Det}~D(\O)}-\fr{3g_j}{m_j(\o_j^2-\O^2)}\Big]\Big)\nonumber
\\
&+&\sum_{j,k=1}^N\fr{mg_jg_k\O^2}{2}{\rm Im}\Big[\fr{g_j g_k m
m_{\rm r}\O^2\o_0^2}{m_j
m_k(\o_j^2-\O^2)(\o_k^2-\O^2)}\alpha_{\rho\rho}(\O)\nonumber \\
&-&\fr{2g_j g_k 
m_{\rm r}\O^4\lambda(\O) e^2G(\O)B^2}{m_j
m_k(\o_j^2-\O^2)(\o_k^2-\O^2)c^2{\rm Det}~D(\O)}
+\fr{3\delta_{jk}}{m_k(\o_k^2-\O^2)}\Big]\nonumber \\
&+&\sum_{j=1}^N\fr{1}{2}m_j(\o_j^2+\O^2){\rm Im}\Big[\fr{(g_j)^2 m
m_{\rm r}\O^2\o_0^2}{(m_j)^2(\o_j^2-\O^2)^2}\alpha_{\rho\rho}(\O)-\fr{2(g_j)^2 
m_{\rm
r}\O^4\lambda(\O)
B^2e^2G(\O)}{(m_j)^2(\o_j^2-\O^2)^2c^2{\rm Det}~D(\O)}\Big]\Big], \nonumber \\
\eea
where we have used Eqs (\ref{betadefinition}), (\ref{gammasymmetric}),
and (\ref{deltasymmetric}). After simplification, we get
\bea
&&U_0(T,B)\nonumber \\
&&=\fr{\hbar}{\pi}\int_0^\infty d\O
\coth\Big[\fr{\hbar\O}{2k_BT}\Big]{\rm Im}\Big[\Big\{\fr{1}{2}m_{\rm r}(\O^2+\o_0^2)+\sum_{j=1}^N\fr{(g_j)^2 m
m_{\rm
r}\O^2\o_0^2(\o_j^2+\O^2)}{2m_j(\o_j^2-\O^2)^2}\Big\}\alpha_{\rho\rho}(\O)\nonumber
\\
&&+\lambda(\O) \Big(\fr{\O
eB}{c}\Big)^2\Big(\fr{d(G(\O))^2}{d\O}\Big)\fr{\o}{2{\rm
Det}~D(\O)}\Big]. 
\l{freeenergysimple}
\eea

From Eqs. (\ref{alphacompact}) and (\ref{DetDomega}), we have
\bea
{\rm Det}~\alpha(\O)&=&[{\rm Det}~D(\O)]^{-1}=\Big[\lambda(\O)\Big[(\lambda(\O))^2-\Big(\fr{\O B
eG(\O)}{c}\Big)^2\Big]\Big]^{-1},
\l{Detalpha}
\eea
while the trace of $\alpha(\O)$ is
\be
\alpha_{\rho \rho}(\O)=\Big[3(\lambda(\O))^2 -
\Big(\fr{\O eG(\O)}{c}\Big)^2B^2\Big]/{\rm Det}~D(\O).
\ee
Following \cite{Li:1990}, we write
\bea
&&\O\fr{d}{d\O}\ln [{\rm Det}~\alpha(\O)]\nonumber \\
&&=-\O\Big\{\fr{d\lambda(\O)}{d\O}\Big[3(\lambda(\O))^2-\Big(\fr{\O B
eG(\O)}{c}\Big)^2\Big]-2\O\lambda(\O)\Big(\fr{BeG(\O)}{c}\Big)^2\Big\}/{\rm Det}~D(\O) \nonumber \\
&&=-3+\Big[\lambda(\O)-\O\fr{d\lambda(\O)}{d\O}\Big]\alpha_{\rho\rho}(\O).
\l{final1}
\eea
From Eq. (\ref{lambda}), we get
\bea
\lambda(\O)-\O\fr{d\lambda(\O)}{d\O}=m_{\rm r}(\O^2+\o_0^2)+\sum_{j=1}^N\fr{(g_j)^2 m
m_{\rm
r}\O^2\o_0^2(\o_j^2+\O^2)}{m_j(\o_j^2-\O^2)^2}.
\l{final2}
\eea
Now, using Eqs. (\ref{Detalpha}), (\ref{final1}) and (\ref{final2}) in Eq.
(\ref{freeenergysimple}), we get
\bea
U_0(T,B)&=&\fr{1}{\pi}\int_0^\infty d\O~ u(\O,T)
{\rm Im}\Big[\fr{d}{d\O}\ln [{\rm
Det}~\alpha(\O)]+\lambda(\O) \Big(\fr{\O
eB}{c}\Big)^2\Big(\fr{d(G(\O))^2}{d\O}\Big){\rm
Det}~\alpha(\O)\Big], \nonumber \\ 
\l{energyfinal}
\eea
where $u(\O,T)$ is the Planck energy of a free oscillator of frequency
$\O$:
\be
u(\O,T)=\fr{\hbar \O}{2}\coth\Big[\fr{\hbar\O}{2k_BT}\Big].
\ee
\subsection{Free energy}
The free energy $F_0(T,B)$ of the charged particle is obtained from Eq.
(\ref{energyfinal}) for the mean internal energy, since the two quantities are related as
$U_0(T,B)=F_0(T,B)-T\fr{\partial F_0(T,B)}{\partial T}$. We get
\bea
F_0(T,B)&=&\fr{1}{\pi}\int_0^\infty d\O~ f(\O,T){\rm Im}\Big[\fr{d}{d\O}\ln [{\rm
Det}~\alpha(\O)]\nonumber \\
&+&\lambda(\O) \Big(\fr{\O
eB}{c}\Big)^2\Big(\fr{d(G(\O))^2}{d\O}\Big){\rm
Det}~\alpha(\O)\Big],  
\l{freeenergyfinal}
\eea
where $f(\O,T)$ is the free energy of a free oscillator of frequency
$\O$:
\be
f(\O,T)=k_BT \ln \Big[2 \sinh\Big[\fr{\hbar\O}{2k_BT}\Big]\Big]=\fr{\hbar \o}{2}+k_BT\ln\Big[1-\exp\Big(\fr{-\hbar
\o}{k_BT}\Big)\Big],
\l{f}
\ee
where in the last equality we have separated out the contribution from the zero-point energy.

Equation (\ref{freeenergyfinal}) is the central result of the paper. To
make explicit the contribution of the external magnetic field to the
free energy, we write Eq. (\ref{Detalpha}) as
\bea
{\rm Det}~\alpha(\O)&=&[\alpha^{(0)}(\O)]^3\Big[1-\Big(\fr{\O B
eG(\O)}{c}\Big)^2[\alpha^{(0)}(\O)]^2\Big]^{-1}, 
\l{Detalpha1}
\eea
where
\be
\alpha^{(0)}(\O)=1/\lambda(\O)
\ee
is the susceptibility in the absence of the magnetic field.
Using Eq. (\ref{Detalpha1}) in Eq. (\ref{freeenergyfinal}), we get 
\be
F_0(T,B)=F_0(T,0)+\Delta_1 F_0(T,B)+\Delta_2 F_0(T,B), 
\l{freeenergydecomp}
\ee
where
\bea
F_0(T,0)=\fr{3}{\pi}\int_0^\infty d\O~ f(\O,T){\rm
Im}\Big[\fr{d}{d\O}\ln \alpha^{(0)}(\O)\Big]  
\l{freeenergynofield}
\eea
is the free energy of the charged particle in the absence of the
magnetic field. The contribution from the field is contained in the two 
terms $\Delta_1 F_0(T,B)$ and $\Delta_2 F_0(T,B)$, given by
\bea
\Delta_1 F_0(T,B)&=&-\fr{1}{\pi}\int_0^\infty d\O~ f(\O,T){\rm
Im}\Big[\fr{d}{d\O}\ln \Big\{1-(G(\O))^2\Big(\fr{\O B
e}{c}\Big)^2[\alpha^{(0)}(\O)]^2\Big\}\Big], \nonumber \\ 
\l{freeenergyfinal1}
\eea
and
\bea
\Delta_2 F_0(T,B)&=&\fr{1}{\pi}\int_0^\infty d\O~ f(\O,T){\rm Im}\Big[[\alpha^{(0)}(\O)]^2\Big(\fr{\O
eB}{c}\Big)^2\nonumber \\
&\times&\Big(\fr{d(G(\O))^2}{d\O}\Big)\Big\{1-\Big(\fr{\O B
eG(\O)}{c}\Big)^2[\alpha^{(0)}(\O)]^2\Big\}^{-1}\Big].  
\l{freeenergyfinal2}
\eea

Let us now comment on the form of the free energy
(\ref{freeenergydecomp}) with respect to that for 
coordinate-coordinate coupling between the particle and the heat-bath
oscillators, obtained in Ref. \cite{Li:1990}. In the latter case, the free
energy is given by 
\be
F_0(T,B)=F_0(T,0)+\Delta_1 F_0(T,B), 
\l{freeenergydecomp-cc}
\ee
where $F_0(T,0)$ has the same form as in Eq. (\ref{freeenergynofield}),
while $\Delta_1 F_0(T,B)$ is given by
\bea
&&\Delta_1 F_0(T,B)=-\fr{1}{\pi}\int_0^\infty d\O~ f(\O,T){\rm Im}\Big[\fr{d}{d\O}\ln \Big\{1-\Big(\fr{\O B
e}{c}\Big)^2[\alpha^{(0)}(\O)]^2\Big\}\Big]. 
\l{freeenergyfinal-cc}
\eea
We thus see that with respect to coordinate-coordinate coupling, the
free energy has two differences, namely, (i) the appearance of the extra factor
$(G(\O))^2$ in the term $\Delta_1 F_0(T,B)$, and 
(ii) the presence of the additional term $\Delta_2 F_0(T,B)$. 
With respect to the two schemes of coupling, the differences observed here
in the form of the thermodynamic potential, i.e., the free energy, add to
the ones noted in the reduced dynamical
description of the charged particle by means of the quantum Langevin
equation \cite{Gupta:2011}.
   
\section{Explicit free energy for an illustrative heat-bath spectrum}
In this section, we utilize the results obtained in the previous section
to obtain and analyze the low-temperature thermodynamic behavior of the
charged particle. To proceed, we first express the integrands in Eqs.
(\ref{freeenergynofield}), (\ref{freeenergyfinal1}), and
(\ref{freeenergyfinal2}) in terms of the Laplace transform of the diagonal part of the memory
function, given by \cite{Gupta:2011}
\be
\widetilde{\mu}_{\rm d}(\o)=i\sum_{j=1}^N\fr{g_j^2mm_{\rm
r}\o_0^2\o}{m_j(w^2-\o_j^2)}; 
\ee
one has
\be
\alpha^{(0)}(\o)=\fr{1}{m_{\rm r}(\o_0^2-\o^2)-i\o\widetilde{\mu}_{\rm
d}(\o)},
\ee
and
\be
G(\o)=\fr{m_{\rm r}}{m}-\fr{i\o\widetilde{\mu}_{\rm d}(\o)}{m\o_0^2},
\ee
so that we get
\be
 F_0(T,0)=\fr{3}{\pi}\int_0^\infty d\O~ f(\O,T)I_1,
\l{F0}
\ee
where
\bea
&&I_1={\rm Im}\Big[\fr{d}{d\omega}\ln \alpha^{(0)}(\o)\Big]=\fr{m_{\rm
r} \Big((\o^2 + \o_0^2) \widetilde{\mu}_d(\o) + \o (-\o^2 +
\o_0^2)\fr{d \widetilde{\mu}_d(\o)}{d\o}\Big)}{m_{\rm r}^2 (\o^2 - \o_0^2)^2 + \o^2
\widetilde{\mu}_{\rm d}^2(\o)}.
\l{I1}
\eea
Also, we have
\be
\Delta_1 F_0(T,B)=-\fr{1}{\pi}\int_0^\infty d\O~ f(\O,T)I_2,
\l{Delta1}
\ee
where
\be
I_2={\rm Im}\Big[\fr{d}{d\o}\ln
\Big\{1-(G(\O))^2\Big(\fr{eB\o}{c}\Big)^2[\alpha^{(0)}(\o)]^2\Big\}\Big]=\fr{A}{B};
\l{I2}
\ee
\bea
A&=&2 m_{\rm r} \o^4 \o_c^2 \Big[-m_{\rm r}^4 \o_0^6 \Big(\o^6 + 5
\o_0^6 + \o^4 (3 \o_0^2 + \o_c^2) - 3 \o^2 \o_0^2 (3 \o_0^2 +
\o_c^2)\Big) \widetilde{\mu}_d(\o) \nonumber \\
&+& 
m_{\rm r}^2 \o^2 \o_0^2 \Big(-8 \o_0^6 + \o^4 (\o_0^2 - \o_c^2) + \o^2
(9 \o_0^4 + 4 \o_0^2 \o_c^2)\Big) \widetilde{\mu}_{\rm d}^3(\o)\nonumber
\\
&+& (-3
\o^4 \o_0^4 + \o^6 \o_c^2) \widetilde{\mu}_{\rm d}^5(\o) \nonumber \\
&+&m_{\rm r}^4 \o \o_0^6 (\o^2 - \o_0^2) \Big(\o^4 + \o_0^4 - \o^2 (2
    \o_0^2 + \o_c^2)\Big)\fr{d \widetilde{\mu}_{\rm d}(\o)}{d\o} \nonumber \\
    &-& 
    m_{\rm r}^2 \o^5 \o_0^2 \Big(-3 \o_0^4 + \o^2 (3 \o_0^2 +
    \o_c^2)\Big)
    \widetilde{\mu}_{\rm d}^2(\o)\fr{d \widetilde{\mu}_{\rm d}(\o)}{d\o}  + \o^5
    (\o_0^4 - \o^2
    \o_c^2)\widetilde{\mu}_{\rm d}^4(\o)\fr{d \widetilde{\mu}_{\rm
    d}(\o)}{d\o}\Big],
    \nonumber \\
\l{A}
\eea
\bea
B&=&\Big(m_{\rm r}^2 (\o^2 - \o_0^2)^2 + \o^2 \widetilde{\mu}_{\rm
d}^2(\o)\Big)\Big[m_{\rm r}^4 \o_0^8 \Big(\o^4 + \o_0^4 - \o^2 (2 \o_0^2 +
\o_c^2)\Big)^2 \nonumber \\
&+& 
    2 m_{\rm r}^2 \o^2 \o_0^4 \Big(\o_0^8 + \o^6 \o_c^2 - 2 \o^2 \o_0^4
    (\o_0^2 + \o_c^2) +  \o^4 (\o_0^2 + \o_c^2)^2\Big)
    \widetilde{\mu}_{\rm d}^2(\o) \nonumber \\
    &+& \o^4 (\o_0^4 - \o^2 \o_c^2)^2
    \widetilde{\mu}_{\rm d}^4(\o)\Big],
\l{B}
\eea
and
\be
\o_c=\fr{eB}{mc}
\l{wc}
\ee
is the cyclotron frequency.
Similarly, Eq. (\ref{freeenergyfinal2}) may be rewritten as
\be
\Delta_2 F_0(T,B)=\fr{1}{\pi}\int_0^\infty d\O~ f(\O,T)I_3,
\l{Delta2}
\ee
where
\bea
I_3&=&{\rm Im}\Big[[\alpha^{(0)}(\O)]^2\Big(\fr{\O
eB}{c}\Big)^2\Big(\fr{d(G(\O))^2}{d\O}\Big)\Big\{1-\Big(\fr{\O B
eG(\O)}{c}\Big)^2[\alpha^{(0)}(\O)]^2\Big\}^{-1}\Big]\nonumber \\ 
&=&\Big[-2 m_{\rm r} \o^2 \o_0^2 \o_c^2 \Big\{m_{\rm r}^2 \o_0^4 \Big(\o^4 +
\o_0^4 - \o^2 (2 \o_0^2 + \o_c^2)\Big)+\o^2 \Big(\o_0^4 - \o^2 (2 \o_0^2 +
\o_c^2)\Big) \widetilde{\mu}_{\rm d}^2(\o)\Big\}\nonumber \\
&\times&\Big(\widetilde{\mu}_{\rm d}(\o)+
\o\fr{d \widetilde{\mu}_{\rm d}(\o)}{d\o}\Big)\Big]\Big[m_{\rm r}^4 \o_0^8 \Big(\o^4 + \o_0^4 -
\o^2 (2 \o_0^2 + \o_c^2)\Big)^2 \nonumber \\
&+& 
  2 m_{\rm r}^2 \o^2 \o_0^4 \Big(\o_0^8 + \o^6 \o_c^2 - 2 \o^2 \o_0^4
  (\o_0^2 + \o_c^2) + 
     \o^4 (\o_0^2 + \o_c^2)^2\Big) \widetilde{\mu}_{\rm
     d}^2(\o)\nonumber \\
     &+& \o^4
     (\o_0^4 - \o^2 \o_c^2)^2 \widetilde{\mu}_{\rm d}^4(\o)\Big]^{-1}.
\l{I3}
\eea
One can check that $I_1$, $I_2$, and $I_3$ all have the dimensions of $1/\o$ as
required.

Now, $f(\O,T)$ vanishes exponentially for frequencies $\O \gg
k_BT/\hbar$. Then, in order to evaluate the free energy of the charged particle
at low temperatures, we need to consider only low-$\O$ contributions in evaluating the
integrals in Eqs. (\ref{F0}), (\ref{Delta1}) and (\ref{Delta2}). To
proceed, let us consider a heat-bath for which the memory function for
small $\O$ has the form
\be
\widetilde{\mu}_{\rm d}(\o)=m_{\rm r}b^{1-\nu}(-i\o)^{\nu};    ~~-1<\nu <1,
\l{mud}
\ee
where $b$ is a positive constant with the dimension of frequency and
$\nu$ is within the indicated range so that $\widetilde{\mu}_{\rm d}(\o)$ is a
positive real function, as required \cite{Ford:1988,Ford:2005}. The Ohmic, sub-Ohmic and super-Ohmic heat-bath
spectra are obtained by considering $\nu=0$, $-1<\nu<0$, and
$0 < \nu<1$, respectively. Using Eq. (\ref{mud}) in Eqs. (\ref{I1}),
(\ref{I2}) and (\ref{I3}), we
find that to lowest order in $\O$, one has
\be
I_1= \fr{(1+\nu)b^{1-\nu}}{\O_0^2}\cos \Big(\fr{\nu
\pi}{2}\Big)\O^{\nu},
\ee
while to the same order in $\o$, the quantities $I_2$ and $I_3$ have
no contributions. We use the above expression for $I_1$ in Eq.
(\ref{F0}), and the result 
\be
\int_0^\infty dy ~y^\nu \log(1- e^{-y})=-\Gamma(\nu+1)\zeta(\nu+2),
\ee
where $\Gamma(z)$ is the gamma function, while $\zeta(z)$ is the Riemann
 zeta-function,
\be
\zeta(z)=\sum_{n=1}^\infty \fr{1}{n^z},
\ee
to finally get that at low temperatures, 
\bea
F_0(T,B)=F_0(T,0)=-3\Gamma(\nu+2)\zeta(\nu+2)\cos\Big(\fr{\nu\pi}{2}\Big)\fr{\hbar
b^3}{\pi\O_0^2}\Big(\fr{k_BT}{\hbar b}\Big)^{\nu+2}.
\eea
It is readily seen that the entropy $S=-\fr{\partial F}{\partial T}$ approaches zero as
$T\rightarrow 0$, in agreement with the third law of thermodynamics
\cite{Nernst:1911}.
\section{Conclusions}
\l{conclusions}
In this paper, we derived an exact formula for the equilibrium free
energy of a charged quantum particle moving in a harmonic potential in the
presence of a uniform magnetic field and
linearly coupled to a heat bath of independent quantum harmonic
oscillators through the momentum variables. The free energy has an
expression which is different from that for the case of 
coordinate-coordinate coupling between the particle and the heat-bath
oscillators. For an
illustrative heat-bath spectrum, we evaluated the free
energy for non-zero magnetic fields in the low-temperature limit, showing thereby that the entropy of the
charged particle vanishes at zero temperature, in conformity with the
third law of thermodynamics.
\section{Acknowledgments}
S. G. acknowledges support of the Israel Science Foundation
and the French contract ANR-10-CEXC-010-01. We acknowledge useful
correspondence with Alessandro Cuccoli.

\section*{References}

\end{document}